\documentclass[journal]{IEEEtran}

\usepackage{float}
\usepackage{cite}									
\usepackage{amsmath}									
\usepackage{graphicx}
\usepackage{epstopdf}
\usepackage{enumitem}
\usepackage[cmyk]{xcolor}
\usepackage{pgfplots}
\pgfplotsset{compat=1.5}
\usepackage[caption=false]{subfig}
\allowdisplaybreaks
\emergencystretch=\maxdimen
\pgfplotsset{legend image with text/.style={legend image code/.code={%
\node[anchor=center] at (0.3cm,0cm) {#1};}},}

\usepackage[keeplastbox]{flushend} 

\begin{document}

\title{5G Mobile Cellular Networks: Enabling Distributed State Estimation for Smart Grids}

\author{Mirsad Cosovic,
		Achilleas Tsitsimelis,
        Dejan Vukobratovic,
        Javier Matamoros,
       	Carles Ant\'on-Haro% <-this % stops a space
       	
\thanks{M. Cosovic is with Schneider Electric DMS NS, Novi Sad, Serbia (e-mail: mirsad.cosovic@schneider-electric-dms.com).  A. Tsitsimelis, J. Matamoros and C. Ant\'on-Haro are with Centre Tecnol\`ogic de Telecomunicacions de Catalunya (CTTC/iCERCA), Castelldefels - Barcelona, Spain (e-mail: \{achilleas.tsitsimelis, javier.matamoros, carles.anton\}@cttc.es). D. Vukobratovic is with Department of Power, Electronic and Communications Engineering, University of Novi Sad, Novi Sad, Serbia (e-mail: dejanv@uns.ac.rs).} }

\maketitle

\begin{abstract}
With transition towards 5G, mobile cellular networks are evolving into a powerful platform for ubiquitous large-scale information acquisition, communication, storage and processing. 5G will provide suitable services for mission-critical and real-time applications such as the ones envisioned in future Smart Grids. In this work, we show how emerging 5G mobile cellular network, with its evolution of Machine-Type Communications and the concept of Mobile Edge Computing, provides an adequate environment for distributed monitoring and control tasks in Smart Grids. In particular, we present in detail how Smart Grids could benefit from advanced distributed State Estimation methods placed within 5G environment. We present an overview of emerging distributed State Estimation solutions, focusing on those based on distributed optimization and probabilistic graphical models, and investigate their integration as part of the future 5G Smart Grid services.
\end{abstract}

\IEEEpeerreviewmaketitle
\vspace{-0.35cm}
\section{Introduction}

In recent years, two main trends have emerged in the evolution of power grids: i) the de-regulation of energy markets, and ii) the increasing penetration of renewable energy sources. The former results in an increased exchange of large amounts of power between adjacent areas, possibly under the control of different regional utilities. The latter leads to larger system dynamics, due to the intermittency of renewable energy sources. To ensure power grid stability, such variations must be timely and accurately monitored.
  
State Estimation (SE) is a key functionality of electric power grid's energy management systems. SE aims to provide an estimate of the system state variables (voltage magnitude and angles) at \emph{all} the buses of the electrical network from a set of remotely acquired measurements. The centralized (classical) SE schemes may prove inapplicable to emerging decentralized and dynamic power grids, due to large communication delays and high computational complexity that compromise their ability for real-time operation. Hence, the interest of the community is shifting from centralized to distributed SE algorithms based on more sophisticated optimization techniques beyond the classical iterative Gauss-Newton approaches \cite{exposito}. 

Instrumental to this evolution is the deployment of synchronized Phasor Measurement Units (PMUs) able to accurately measure voltage and current phasors at high sampling rates. Exploiting PMU inputs by robust, decentralized and real-time SE solution calls for novel communication infrastructure that would support future Wide Area Monitoring System (WAMS). WAMS aims to detect and counteract power grid disturbances in \emph{real time}, thus requiring a communication infrastructure able to: i) integrate PMU devices with extreme reliability and ultra-low (millisecond) latency, ii) provide support for distributed and real-time computation architecture for future SE algorithms, and iii) provide backwards compatibility to legacy measurements traditionally collected by Supervisory Control and Data Acquisition (SCADA) systems.

The advent of 5G communication networks will largely facilitate the provision of the distributed information acquisition and processing services required in WAMS systems. As far as information \emph{acquisition} is concerned, the introduction of massive Machine-Type Communication (mMTC) services will allow for a large-scale deployment Advanced Metering Infrastructure (AMI). For those measurement devices (e.g. PMUs) requiring both very low latencies and very high reliabilities, resorting to Ultra-Reliable Low-Latency Communications (URLLC) services \cite{ghosh} will be needed. As for information \emph{processing}, novel architectural concepts such as Mobile Edge Computing (MEC) will be key for the deployment of the aforementioned distributed SE approaches \cite{young}. 

The purpose of this work is twofold: i) to discuss the fundamental role to be played by 5G networks as an enabler of advanced distributed SE schemes, and ii) to place two promising distributed SE solutions (based on distributed optimization and probabilistic graphical models) in such a 5G communications scenario. Specifically, first we describe how distributed SE can be integrated into the framework of MEC, while acquiring measurements via 5G Machine-Type Communications (MTC) services. Then, we focus on the two distributed SE approaches which are based either on the Alternating Direction Method of Multipliers (ADMM), or on probabilistic graphical models and Belief Propagation (BP) algorithms. We also assess their performance and discuss their applicability in realistic 5G communication scenarios, using the corresponding \emph{centralized} versions of the SE schemes as benchmarks.

\section{5G Enhancements for Distributed Information Acquisition and Processing in Smart Grids}

In this section, we review 5G radio interface enhancements and MEC concepts as the main enablers for future 5G smart grid applications.

\subsection{Radio Interface Enhancements for 5G}

3GPP standards providing radio interface enhancements for MTC have been recently adopted within 3GPP LTE Release 13. Three solutions for MTC services are introduced: enhanced Machine-Type Communications (eMTC), Narrow-Band Internet of Things (NB-IoT), and Extended Coverage GSM Internet of Things (EC-GSM-IoT) \cite{yavuz}. For legacy (SCADA) measurement devices such as Remote Terminal Units (RTUs), suitable solution is provided by eMTC albeit with the same reliability and latency guarantees as provided by the LTE physical layer (PHY). In contrast, for massive access of low-rate low-cost devices, new NB-IoT or EC-GSM-IoT extensions provide proper solution. For example, NB-IoT targets up to $50\mathrm{K}$ devices per macro-cell with extended coverage, thus providing ideal solution for smart meter data acquisition. However, significant improvements over 4G radio interface both in reliability and latency are needed for real-time services relying on PMUs. 

3GPP 5G standardization of the New Radio (NR) interface is initiated in Release 13 with a requirement and architecture study. Following ITU 5G requirements, NR will support two new services suitable for machine-type devices: massive Machine-Type Communications and Ultra-Reliable Low-Latency Communications. mMTC will further enhance massive connectivity provisioning established by NB-IoT, targeting connection density of $10^6\,\mathrm{devices/km^2}$ in urban dense scenario while offering Packet Loss Rates (PLRs) below $1 \%$. For URLLC, the generic radio interface latency target is $0.5\,\mathrm{ms}$, while reliability targets PLR of $10^{-5}$ for $32\,\mathrm{byte}$ packets and $1\,\mathrm{ms}$ latency. As we detail later, URLLC represents suitable solution for WAMS real-time services that target system monitoring and control at the PMU sampling rates. However, in order to meet the stringent URLLC requirements, not only radio interface, but also the core network architecture will require novel solutions. 

\subsection{Mobile Edge Computing in 5G}
 
The centralization and virtualization of core network functions within the so called Cloud RAN (C-RAN) architecture reduces costs and complexity of Radio Access Network (RAN) densification. Further, the adoption of Mobile Cloud Computing (MCC) architectures, allowing user equipment to offload computation and store data in remote cloud servers, facilitates the deployment of a number of novel user application and services. However, the major drawback associated to the MCC architecture is the large latency between the end user and the remote cloud center, thus limiting the applicability of MCC for services requiring very low latencies. This has led to a recent surge of interest in MEC architectures, where cloud computing and storage is distributed and pushed towards the mobile network edge \cite{young}. With MEC\footnote{With a slight abuse of notation, hereinafter we use the term 'MEC'. In some passages, however, 'fog computing' could be deemed to be more appropriate.}, many applications and services will benefit from localized communication, storage, processing and management, thus dramatically decreasing service response latency, reducing the traffic load on the core network, and improving context-awareness \cite{addepalli}. MEC concept is not in collision with MCC; they complement each other in building flexible and reconfigurable 5G networks using a ''network slicing'' approach, where different services may be easily instantiated using different virtualized architectures on top of the high-performance MEC host nodes. Instrumental to the development of flexible packet core networks are novel 5G reconfigurable architectures based on Software Defined Networking (SDN) and Network Function Virtualization (NFV) \cite{turtinen}. Thus, enhanced with support for MEC, 5G mobile core networks will provide a distributed information processing and storage architecture that is ideally suited for services requiring low-latency and localized decision making.    
 
\section{Distributed State Estimation Methods}

The deregulation of energy markets, necessity for real-time monitoring and control, along with utilities' data security and privacy concerns in multi-area settings substantiate the need for developing \emph{hierarchical} and \emph{distributed} SE methods as an alternative to classical \emph{centralized} schemes.

\subsection{Hierarchical and Distributed SE Methods}

In \emph{hierarchical} SE \cite{exposito}, \cite{korres}, a central authority controls the local processor in each area or level. Gomez-Exposito et al. \cite{exposito} propose a hierarchical multi-level SE scheme where local estimates are computed at lower voltage levels and transferred to higher voltage areas, up to the system operator level, in order to estimate the system-wide state. In each stage, the SE problem is solved via the Gauss-Newton method. Still in a hierarchical context, Korres in \cite{korres} proposes to decompose, on a geographical basis, the overall system into a number of subsystems in non-overlapping areas. Each area independently runs its own gradient-based SE scheme on the basis of local measurements. Such estimates are then communicated to the central coordinator which computes the system-wide solution.
 
In \emph{distributed} approaches \cite{minguez, giannakis, haro}, on the contrary, each local processor communicates only with its neighbors, since no central authority exists. The authors in \cite{minguez} propose a distributed SE scheme based on primal-dual decomposition. This method requires the exchange of information only between \emph{neighboring} areas, namely, border state variables and the dual variables. For each area, the problem is solved through classical Gauss-Newton techniques. Differently, Kekatos and Giannakis \cite{giannakis} resort the ADMM to solve the SE problem in a distributed fashion. In contrast to \cite{minguez}, the authors develop a robust version leveraging on the sparsity of bad data measurements. Going one step beyond, \cite{zhu} proposes a \emph{hybrid} scheme including \emph{both} PMUs and legacy measurements. Here, the SE problem is casted into a semidefinite programming framework and solved via convex semidefinite relaxation techniques, both in centralized and decentralized settings. In \cite{haro}, the authors propose a hybrid multi-area state estimator based on successive convex approximation (SCA) and ADMM. The proposed distributed approach is equivalent to the centralized case in terms of estimation accuracy and is able to operate in broader scenarios where the semidefinite relaxation approach fails. 

Going one step beyond, other authors \cite{kavcic}, \cite{vukobratovic}, have considered \emph{fully distributed} SE approaches where interaction takes place at the bus level rather than the area level.
Hu et al \cite{kavcic} pioneered in the application of message-passing BP algorithm to the SE problem, where the system state is modeled as a set of stochastic variables. This provides a flexible solution for the inclusion in the model of, e.g., distributed power sources, environmental correlation via historical data and time-varying loads, etc. In a recent work, a distributed Gauss-Newton algorithm based on factor graphs and BP algorithm is proposed and shown to provide the same accuracy as the centralized Gauss-Newton algorithm \cite{vukobratovic}, while being flexible enough to accommodate both fully distributed and multi-area SE scenario. 

As a representative methods, ADMM and BP are particularly promising and, thus, they will be described with further detail later in this section. Prior to that, we describe a system model suitable for both approaches.

\subsection{System Model}

The SE aims to determine the values of the state variables based on the knowledge of the network topology and measurements obtained from devices spread across the power system. Thus, the SE problem reduces to solving the system of equations: $\mathbf{z}=$ $\mathbf{h}(\mathbf{x})+$ $\mathbf{u}$, where $\mathbf{h}(\mathbf{x})=$ $(h_1(\mathbf{x}),$ $\dots,$ $h_k(\mathbf{x}))$ may include both non-linear (from legacy metering devices) and linear measurement functions (from PMUs); $\mathbf{x}=$ $(x_1,$ $\dots,$ $x_n)$ is the vector of the state variables; $\mathbf{z}=$ $(z_1,$ $\dots,$ $z_k)$ is the vector of independent measurements (where $n < k$), and $\mathbf{u}=$ $(u_1,$ $\dots,$ $u_k)$ is the vector of measurement errors. The state variables are bus voltage magnitudes and bus voltage angles, along with transformer magnitudes of turns ratio and transformer angles of turns ratio. Figure \ref{Fig_ieee30} below illustrates a possible scenario for the collection of measurements in the IEEE 30 bus test case. 
	\begin{figure}[ht]
	\centering
	\includegraphics[width=8.5cm]{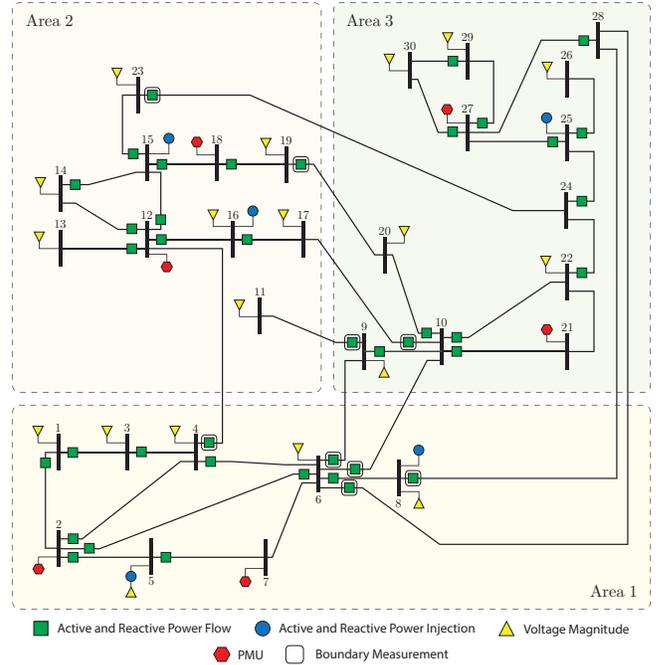}
	\caption{The IEEE 30 bus test case segmented into three areas 
	with a given collection of measurements.}
	\label{Fig_ieee30}
	\end{figure}

\subsection{Optimization-Based Distributed SE Methods}

The ADMM is experiencing renewed popularity after its discovery in the mid-twentieth century. ADMM was conceived to overcome the weaknesses of its predecessors: the primal-dual decomposition method and the method of multipliers. The former is suitable for distributed optimization but presents convergence issues for non-differentiable objectives. Conversely, the method of multipliers can deal with non-smooth functions but couples the objective function which makes it barely suitable for distributed optimization. The ADMM brings the two features together: it is suitable for distributed implementation and can efficiently deal with non-differentiable objective functions. 

The canonical optimization problem solved by ADMM is the minimization of a composite objective function, i.e., $f(\mathbf x)+$ $g(\mathbf z)$, subject to a linear equality constraint of the form $\mathbf {Ax}+$ $\mathbf {Bz} = $ $\mathbf {c}$, with $\mathbf x$ and $\mathbf z$ being the optimization variables.  To deal with non-differentiable functions, the ADMM augments the cost function by a quadratic penalty term that transforms the optimization problem into a strongly convex problem but with the same stationary solution. This transformation has major implications in the dual domain, as the dual function becomes differentiable. Then, the ADMM iterates sequentially on the primal optimization variables, $\mathbf x$ and $\mathbf z$, and the dual variables, until convergence. 
%The major difference with respect to the method of multipliers is the sequential optimization on the primal domain, which opens new ways to decentralize many optimization problems.

To decentralize an optimization problem, ADMM decouples the objective function with consensus variables. Consensus variables introduce equality constraints into the optimization problem, separating it into a number of subproblems \cite{giannakis}, \cite{haro}. The resulting ADMM algorithm can be interpreted as an iterative message passing procedure, in which the agents solving the subproblems (e.g., utilities in the multi-area SE problem) exchange consensus and dual variables until convergence.  Besides, ADMM can be used in conjunction with SCA approaches that efficiently deal with non-convex problems. 
%By doing so, it becomes a versatile and powerful tool to solve many realistic engineering problems.

\subsection{Probabilistic Inference-Based Distributed SE Methods}  

Probabilistic graphical models, such as factor graphs, provide a convenient framework to represent dependencies among the system of random variables, such as the state variables $\mathbf x$ of the power system. Specifically, the bus/branch model with a given measurement configuration is mapped onto an equivalent factor graph containing the set of \emph{factor} and \emph{variable} nodes. \emph{Factor nodes} are defined by the set of measurements: arbitrary factor node $f$ is associated with measured value $z$, measurement error $u$ and measurement function $h(\mathbf x)$. \emph{Variable nodes} are determined by the set of state variables $\mathbf x$. A factor node is connected by an edge to a variable node, if and only if the state variable is an argument of the corresponding measurement function $h(\mathbf x)$.

When applied on factor graphs, BP algorithms allow to efficiently calculate marginal distributions of the system of random variables. BP is a distributed message-passing algorithm in which two types of messages are exchanged along the edges of the factor graph: messages from a variable node to a factor node, and vice-versa. In general, for SE scenarios a \emph{loopy} BP algorithm must be used since the corresponding factor graph contains cycles. Loopy BP is an iterative algorithm in which, for the standard scheduling, messages are updated in parallel in respective half-iterations.  Within half-iterations, factor nodes calculate and send messages to incident variable nodes, while subsequently, variable nodes calculate and send messages back towards factor nodes. As a general rule, an output message on any edge exclusively depends on incoming messages from all other edges. BP messages represent beliefs about variable nodes, thus a message that arrives or departs from a certain variable node is a function (distribution) of the random variable corresponding to the variable node. Finally, the marginal inference provides an estimate of the state variables (voltages in the power system).

\section{Proposed 5G System Architecture for Distributed Smart Grid Services}

In this section, we propose to leverage an emerging 5G network architecture, in particular, its features outlined in Section II, to enable deployment of advanced Smart Grid services such as distributed SE. We also discuss the latency and reliability requirements that distributed SE imposes on 5G communication networks.

	\begin{figure*}[ht]
	\centering
	\includegraphics[width=17.7cm]{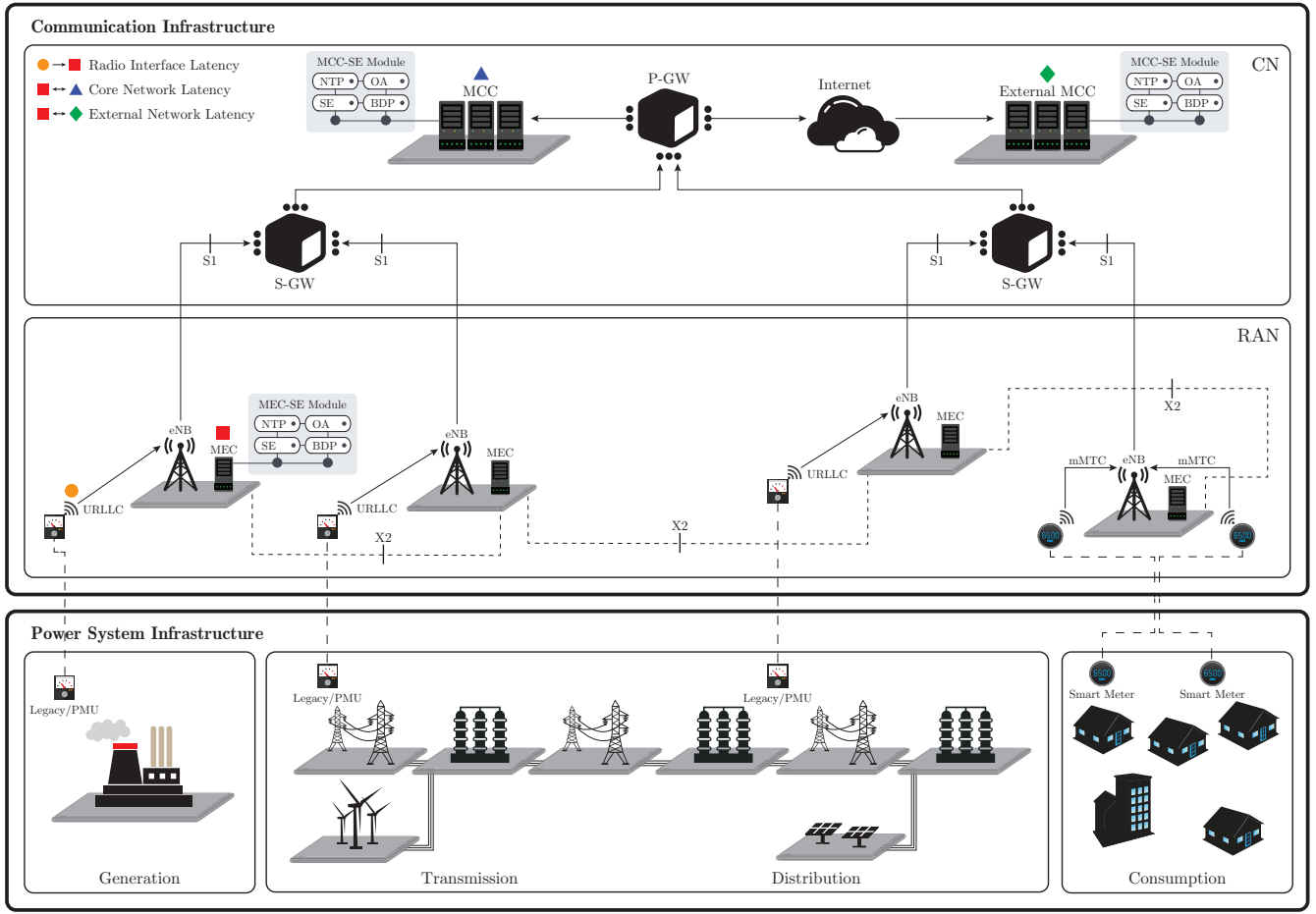}
	\caption{The architecture with two layers: i) power system infrastructure and 
	ii) communication infrastructure that combines novel RAN interfaces supporting 
	mMTC and URLLC, and new virtualized core network (CN) MEC/MCC-based 
	architecture with network topology processor (NTP), observability analysis (OA), 
	state estimation algorithm (SE) and bad data processing (BDP) routines, 
	to support future Smart Grid services such as distributed SE.}
	\label{Fig_archi}
	\end{figure*}

\subsection{System Architecture}

In Figure \ref{Fig_archi}, we present the proposed system architecture for distributed Smart Grid services. The lowermost layer represents the electrical grid broken down into the generation, transmission, distribution and consumption network segments. The grid is equipped with a large number of measurement devices ranging from legacy RTUs to PMUs and massive-scale smart meter infrastructure. We assume the grid is organized into a number of non-overlapping areas. Such a multi-area SE problem represents an input to the distributed SE algorithms discussed earlier.

As far as the communication technology is concerned, hereinafter we will focus on 5G networks. Still, LTE will be used as a reference, where needed. The electrical grid is covered by the RAN comprising large number of base stations (eNBs). Focusing only on data-plane elements, the packet core network consists of packet gateways (S-GW/P-GW) that inter-connect base stations and provide access to external networks (e.g., the Internet). Base stations connect to core network gateways via S1 interfaces, and may also be directly inter-connected via X2 interfaces.

The support for MCC/MEC within the packet core network is provided in the form of a data center for MCC, and in the form of a large-scale deployment of smaller data and computing centers in the vicinity of base stations at the network edge (for MEC). MEC nodes host the distributed Smart Grid applications (SE, topology processor, etc. see Figure 2). In the sequel, we will focus on distributed SE modules denoted as MEC-SE modules. Using NFV concepts, MEC-SE modules may run within the virtualization environment of MEC nodes, i.e., they can be remotely instantiated, removed and orchestrated using the centralized NFV orchestrator.

Connectivity between remote measurement devices, MEC-SE modules and the MCC-SE module is provided by 5G network.  
For connections between measurement devices and local MEC-SE modules, 5G will offer flexible wireless interfaces for different measurement devices. Massive-scale smart meters will upload their data via mMTC service, while more stringent reliability and latency can be offered to RTUs and PMUs via URLLC service. We assume the smart metering data will be delivered as aggregated measurements using data aggregation units. Data flows between MEC-SE modules can be flexibly established via S1 or X2 interfaces. SDN-based concepts could be applied to, e.g., connect the distributed MEC-SE modules to the central MCC-SE module. This module may or may not participate in the distributed SE process and, also, serve as a central function and data repository interfacing other energy management functions.

\subsection{Latency and Reliability Requirements for Distributed SE}  

Utilizing PMU inputs, WAMS will enable power system operators to monitor their power networks in real time. PMUs track system state variables (phasors) with sampling rates of $10-20\,\mathrm{ms}$. Targeting ever faster reaction to system disturbance, decentralized WAMS architecture employing distributed SE with localized decision making promises minimal system response latency. In the following, we discuss how well 5G reliability and latency targets match the vision of future real-time WAMS.

Consider MEC-SE modules each running an entity of the  distributed SE algorithm whose scope is the surrounding geographic area. Every MEC-SE module continuously updates its local state estimates based on high-rate PMU inputs additionally supported by legacy RTU measurements. Furthermore, neighboring MEC-SE modules exchange messages to execute the distributed SE algorithm, thus further refining their state estimates (see Section III). Clearly, the distributed SE performance critically depends on the latency and reliability of the communication links: i) from a PMU to MEC-SE module, and ii) between two neighboring MEC-SE modules. Note that the latency and reliability of MEC-SE to MCC-SE module communication does not affect distributed SE, but it would affect both the hierarchical and centralized SE.

Figure 3 provides an overview of different 4G/5G radio interface solutions that affect reliability/latency trade-off of the link between a measurement device and a MEC-SE module. From the figure, it is evident that current 4G LTE radio interface imposes reliability/latency trade-off limits that prevent the real-time WAMS goals of tracking the system state with the latencies as low as PMUs sampling rates. As discussed in \cite{wiemann}, and empirically investigated in \cite{rupp}, the LTE PHY interface latency may be decreased to $15-20\,\mathrm{ms}$ in the uplink (due to uplink scheduling requests/grants), and down to $4\,\mathrm{ms}$ in the downlink, if both medium access control (MAC) layer hybrid-ARQ (HARQ) and radio link control (RLC) layer Automatic Repeat reQuest (ARQ) mechanism are avoided, however, with modest PLR $\sim$ $10^{-1}$. The PLR may be gradually decreased down $\sim$ $10^{-5}$ by including up to three HARQ retransmissions, and by using RLC ARQ, for the price of increased latency to $\sim$ $40-60\,\mathrm{ms}$, thus allowing only quasi real-time SE. In contrast, 5G URLLC fits the future real-time WAMS targets with radio interface latency of $\sim$ $1\,\mathrm{ms}$ and PLRs $<10^{-5}$. We note that low-cost interfaces such as NB-IoT/5G mMTC could serve massive AMI connections, as well as the needs of legacy SCADA-based snap-shot SE.
	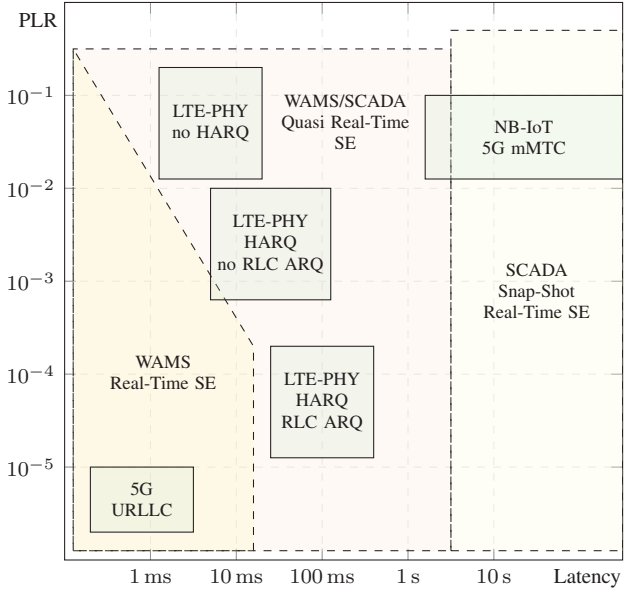
\begin{figure}[ht]
	\centering
	\begin{tikzpicture}
  	\begin{axis}[width=9cm,height=9cm,
  	every axis x label/.style={
    at={(ticklabel* cs:0.86,7.0)},
    anchor=west}, every axis y label/.style={
    at={(ticklabel* cs:0.94,10)},
    anchor=south},
    xlabel={Latency},
    grid=major, grid style={dashed, gray!20}, 
	ylabel={PLR},
    label style={font=\footnotesize},   	
    xtick={1, 2, 3, 4, 5},
    xticklabels={$1\,\mathrm{ms}$, $10\,\mathrm{ms}$, $100\,\mathrm{ms}$, 
    $1\,\mathrm{s}$, $10\,\mathrm{s}$},
    ytick={ 1, 2, 3, 4, 5, 6},
    yticklabels={$10^{-5}$, $10^{-4}$, $10^{-3}$, $10^{-2}$, $10^{-1}$},
    tick label style={font=\footnotesize},
    ymin = 0, ymax = 6,   	
   	xmin = 0, xmax = 6.5]

   	\draw [black, dashed, thin, fill=orange!20, fill opacity=0.2] 
   	(axis cs:0.1,0.1) rectangle (axis cs:4.5,5.5) 
   	node[pos=.5, text width=2.3cm, text=black, 
   	font=\scriptsize, align=center, fill opacity=1, yshift=15ex, xshift=7ex] 
   	{WAMS/SCADA \\ Quasi Real-Time \\ SE};

	\draw[black, dashed, thin, fill=yellow!40, fill opacity=0.2] (axis cs:0.1,0.1) -- 
	(axis cs:0.1,5.5) -- (axis cs:2.2,2.3) -- (axis cs:2.2,0.1) -- cycle
	node[pos=.5, text width=2cm, text=black, 
   	font=\scriptsize, align=center, fill opacity=1, yshift=15ex, xshift=0ex] 
   	{WAMS \\ Real-Time SE};
 
    \draw [black, dashed, thin, fill=yellow!20, fill opacity=0.2] 
   	(axis cs:4.5,0.1) rectangle (axis cs:6.5,5.7) 
   	node[pos=.5, text width=2cm, text=black, 
   	font=\scriptsize, align=center, fill opacity=1, yshift=0ex, xshift=0ex] 
   	{SCADA \\ Snap-Shot Real-Time SE};
   	
   	\draw [black, thin, fill=green!20, fill opacity=0.2] 
   	(axis cs:0.3,0.3) rectangle (axis cs:1.5,1.0) 
   	node[pos=.5, text width=1cm, align=center, text=black, 
   	font=\scriptsize, fill opacity=1] 
   	{5G URLLC};  	
   	   	 	
   	\draw [black, thin, fill=green!20, fill opacity=0.2] 
   	(axis cs:1.7,2.8) rectangle (axis cs:3.1,4) 
   	node[pos=.5, text width=1.7cm, text=black, 
   	font=\scriptsize,align=center, fill opacity=1] 
   	{LTE-PHY HARQ \\ no RLC ARQ};  
      	
   	\draw [black, thin, fill=green!20, fill opacity=0.2] 
   	(axis cs:4.2,4.1) rectangle (axis cs:6.5,5.0) 
   	node[pos=.5, text width=1.5cm, text=black, 
   	font=\scriptsize,align=center, fill opacity=1] 
   	{NB-IoT \\ 5G mMTC}; 
   	   	
   	\draw [black, thin, fill=green!20, fill opacity=0.2] 
   	(axis cs:1.1,4.1) rectangle (axis cs:2.3,5.3) 
   	node[pos=.5, text width=1.1cm, text=black, 
   	font=\scriptsize,align=center, fill opacity=1] 
   	{LTE-PHY no HARQ};   	
   	 
   	\draw [black, thin, fill=green!20, fill opacity=0.2] 
   	(axis cs:2.4,1.1) rectangle (axis cs:3.6,2.3) 
   	node[pos=.5, text width=1.1cm, text=black, 
   	font=\scriptsize,align=center, fill opacity=1] 
   	{LTE-PHY HARQ \\ RLC ARQ};    	 
   	\end{axis}
	\end{tikzpicture}
	\caption{Reliability and latency performance of 4G/5G radio 
	interface solutions relative to different WAMS/SCADA SE services.}
	\label{fig_var}
	\end{figure}

The latency within packet core networks is very low, typically in the range of $3-10\,\mathrm{ms}$ between MEC-SE modules and between MEC-SE to MCC-SE module (for moderately large networks). Note that in the case of two MEC-SE modules residing at two eNBs connected via X2 interface, this latency can be reduced to $\sim$ $1\,\mathrm{ms}$. In contrast, for scenarios where the central MCC-SE module is hosted in an external data center outside the core network, the associated latency can be as high as $10-100\,\mathrm{ms}$. In the following section, we place ADMM and BP-based distributed SE in the context of 5G system architecture, investigating the impact of latency and reliability on the SE performance.

\section{Performance of Distributed State Estimation Methods}

Both BP and ADMM-based distributed SE solutions can be integrated as part of the 5G Smart Grid services described the previous section. For the case of BP, factor graphs of the power system can be flexibly segmented into areas and BP can easily accommodate both intra- and inter-area message exchange, not necessarily with the same periodicity, allowing for asynchronous message scheduling. Thus a factor graph of each area can be maintained within the corresponding MEC-SE module, with local measurements arriving from mMTC and URLLC connections. Inter-area BP messages can be exchanged with a controlled periodicity between the neighboring MEC-SE modules. The exchange of inter-area BP messages establishes a global factor graph and provides the MEC-SE modules with the ability to converge to the global solution. Similarly, for the case of ADMM, the local ADMM-based MEC-SE modules may run single-area optimization based on local topology and measurements. By exchanging messages among neighboring areas, MEC-SE modules iterate through the ADMM optimization process converging towards the global solution.

\subsection{Performance of BP and ADMM-based SE}

In the following, we demonstrate that the state estimate of the distributed BP and ADMM-based algorithms converges to the solution provided by the centralized Gauss-Newton method. We consider both the IEEE 30 and IEEE 118 bus test cases divided into three and nine areas, respectively (see Figure \ref{Fig_ieee30} for IEEE 30 bus test case). The algorithms are tested in three scenarios: i) a measurement configuration with no PMUs, ii) one PMU per area, and iii) two PMUs per area. For a predefined value of the noise variance and using Monte Carlo approach, we generate $500$ random sets of measurement values, feed them to the BP, ADMM and centralized SE algorithms and then compute the average performance. The BP algorithm is implemented as a BP-based distributed Gauss-Newton method described in \cite{vukobratovic}, which can be interpreted as a fully distributed Gauss-Newton method. The ADMM-based algorithm is based on \cite{haro}, where the SCA scheme in the outer loop is combined in a distributed fashion within the iterative framework of ADMM, that constitutes the inner loop. To evaluate both algorithms, we use the Root Mean Square Error (RMSE) after each iteration $k$ $(\mathrm{RMSE}^k)$, normalized by RMSE of the centralized SE algorithm using the Gauss-Newton method after $12$ iterations $(\mathrm{RMSE}_{\mbox{\scriptsize GN}})$. 

\begin{figure}[ht]
	\centering
	\captionsetup[subfigure]{oneside,margin={0.9cm,0cm}}
	\begin{tabular}{@{}c@{}}
	\subfloat[]{\label{Fig_BP1}
	\centering
	\begin{tikzpicture}
  	\begin{semilogyaxis}[width=7.0cm, height=5.5cm,
   	x tick label style={/pgf/number format/.cd,
   	set thousands separator={},fixed},
   	xlabel={Number of iterations $k$},
   	ylabel={$\mathrm{RMSE}_{\mbox{\tiny BP}}^k/\mathrm{RMSE}_{\mbox{\tiny GN}}$},
   	label style={font=\footnotesize},
   	legend columns=2,legend style={legend cell align=left},
   	grid=major,
   	legend style={legend pos=north east,font=\scriptsize},
   	ymin = 0.8, ymax = 150,
   	xmin = 1, xmax = 9,
   	xtick={1,2,3,4,5,6,7,8,9},
   	xticklabels={1,9,36,100,225,441,784,1296,2025},
   	tick label style={font=\footnotesize},
   	ytick={1,10,100}]
	\addlegendimage{legend image with text=IEEE 30}
    \addlegendentry{}
    % column 2 heading
    \addlegendimage{legend image with text=IEEE 118}
    \addlegendentry{}   
    \addplot[mark=diamond*,mark repeat=1, mark size=1.5pt, blue] 
   	table [x={ite}, y={pmu0}] {bp_1_ieee30.txt}; 
   	\addlegendentry{}  
	\addplot[mark=diamond*,mark repeat=1, mark size=1.5pt, blue, dashed] 
   	table [x={ite}, y={pmu0}] {bp_1_ieee118.txt};
   	\addlegendentry{No PMUs}
   	\addplot[mark=otimes*, mark repeat=1, mark size=1.5pt, green]
	table [x={ite}, y={pmu1}] {bp_1_ieee30.txt};
   	\addlegendentry{}
	\addplot[mark=otimes*, mark repeat=1, mark size=1.5pt, green, dashed]
	table [x={ite}, y={pmu1}] {bp_1_ieee118.txt};   	
   	\addlegendentry{1 PMU}
	\addplot[mark=triangle*,mark repeat=1, mark size=1.5pt, red]
   	table [x={ite}, y={pmu2}] {bp_1_ieee30.txt};
   	\addlegendentry{}
   	\addplot[mark=triangle*,mark repeat=1, mark size=1.5pt, red, dashed]
   	table [x={ite}, y={pmu2}] {bp_1_ieee118.txt};
   	\addlegendentry{2 PMUs}
  	\end{semilogyaxis}
 	\end{tikzpicture}}
	\end{tabular}\\
	\begin{tabular}{@{}c@{}}
	\subfloat[]{\label{Fig_ADMM1}
	\begin{tikzpicture}
  	\begin{semilogyaxis}[width=7.0cm, height=5.5cm,
   	x tick label style={/pgf/number format/.cd,
   	set thousands separator={},fixed},
   	xlabel={Number of iterations $k$},
   	ylabel={$\mathrm{RMSE}_{\mbox{\tiny ADMM}}^k/\mathrm{RMSE}_{\mbox{\tiny GN}}$},
   	label style={font=\footnotesize},
   	legend columns=2,legend style={legend cell align=left},
   	grid=major,
   	legend style={legend pos=north east,font=\scriptsize},
   	ymin = 0.8, ymax = 150,
   	xmin = 1, xmax = 3500,
   	xtick={1,500,1000,1500,2000, 2500, 3000, 3500},
   	tick label style={font=\footnotesize},   	
   	ytick={1,10,100}]
	\addlegendimage{legend image with text=IEEE 30}
    \addlegendentry{}
    % column 2 heading
    \addlegendimage{legend image with text=IEEE 118}
    \addlegendentry{}
    \addplot[mark=diamond*,mark repeat=500, mark size=1.5pt, blue] 
   	table [x={ite}, y={pmu0}] {admm_1_ieee30.txt}; 
   	\addlegendentry{}  
	\addplot[mark=diamond*,mark repeat=71, mark size=1.5pt, blue, dashed] 
   	table [x={ite}, y={pmu0}] {admm_1_ieee118.txt};
   	\addlegendentry{No PMUs}
   	\addplot[mark=otimes*, mark repeat=500, mark size=1.5pt, green]
	table [x={ite}, y={pmu1}] {admm_1_ieee30.txt};
   	\addlegendentry{}
	\addplot[mark=otimes*, mark repeat=71, mark size=1.5pt, green, dashed]
	table [x={ite}, y={pmu1}] {admm_1_ieee118.txt};   	
   	\addlegendentry{1 PMU}
	\addplot[mark=triangle*,mark repeat=500, mark size=1.5pt, red]
   	table [x={ite}, y={pmu2}] {admm_1_ieee30.txt};
   	\addlegendentry{}
   	\addplot[mark=triangle*,mark repeat=71, mark size=1.5pt, red, dashed]
   	table [x={ite}, y={pmu2}] {admm_1_ieee118.txt};
   	\addlegendentry{2 PMUs}
  	\end{semilogyaxis}
 	\end{tikzpicture}}
	\end{tabular}
	\caption{Normalized RMSE for the BP algorithm (subfigure a) and ADMM 
	(subfigure b) in three scenarios: without PMUs, one PMU and two PMUs per 
	area for the IEEE 30 and IEEE 118 bus test case.}
	\label{Fig_sce1}
	\end{figure}
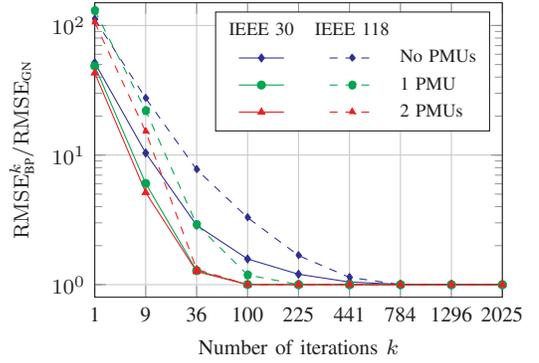
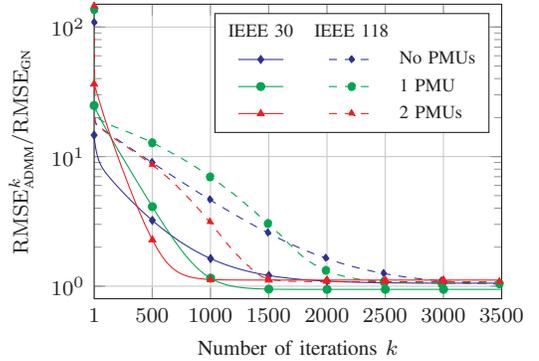

Figure \ref{Fig_BP1} shows that the BP algorithm converges to the solution of the centralized SE for each scenario. As expected, the BP algorithm converges faster for measurement configurations with PMUs. In general, configurations with PMUs can dramatically improve numerical stability of the BP algorithm and prevent oscillatory behavior of messages. Figure \ref{Fig_ADMM1} illustrates the performance of the ADMM-based algorithm. The scheme attains the same performance as the centralized SE. We observe how an increased number of PMUs leads to significant improvement in convergence behavior for both the IEEE 30 and IEEE 118 bus test cases. For the latter, the graph reveals that the algorithm needs a larger number of iterations to converge. Note that, in both cases, non-linear measurement functions (SCADA) are used, and the algorithm is initialized in a (flat-start) state distant from the solution, allowing only snap-shot SE with order of seconds latency. 

\subsection{Distributed SE Methods: Performance vs Reliability}
\begin{figure}[ht]
	\centering
	\begin{tikzpicture}
  	\begin{semilogyaxis}[width=7.2cm, height=5.5cm,
   	x tick label style={/pgf/number format/.cd,
   	set thousands separator={},fixed},
   	label style={font=\footnotesize},
   	xlabel={PLR},
   	ylabel={RMSE},
   	grid=major,
   	legend columns=2,
   	legend style={legend pos=north west,font=\scriptsize, column sep=-0.1cm, 
   	row sep=-0.04cm, legend cell align=left},
   	xmin = 1, xmax = 6,
   	xtick={1,2,3,4,5,6},
   	xticklabels = {$0$,$10^{-5}$,$10^{-4}$,$10^{-3}$,$10^{-2}$,$10^{-1}$},
   	tick label style={font=\footnotesize}]
	\addlegendimage{legend image with text=Gauss-Newton}
    \addlegendentry{}
    % column 2 heading
    \addlegendimage{legend image with text=ADMM}
    \addlegendentry{}
   
    \addplot[mark=square*,mark repeat=1, mark size=1.5pt, blue] 
   	table [x={plr}, y={y6}] {admm_plr.txt}; 
   	\addlegendentry{}  
	\addplot[mark=square*,mark repeat=1, mark size=1.5pt, blue, dashed] 
   	table [x={plr}, y={y3}] {admm_plr.txt};
   	\addlegendentry{$\sigma_\mathrm{pm} = 0.01$}     
   
    \addplot[mark=*,mark repeat=1, mark size=1.5pt, green] 
   	table [x={plr}, y={y5}] {admm_plr.txt}; 
   	\addlegendentry{}  
	\addplot[mark=*,mark repeat=1, mark size=1.5pt, green, dashed] 
   	table [x={plr}, y={y2}] {admm_plr.txt};
   	\addlegendentry{$\sigma_\mathrm{pm} = 0.1$}

    \addplot[mark=diamond*,mark repeat=1, mark size=1.5pt, red] 
   	table [x={plr}, y={y4}] {admm_plr.txt}; 
   	\addlegendentry{}  
	\addplot[mark=diamond*,mark repeat=1, mark size=1.5pt, red, dashed] 
   	table [x={plr}, y={y1}] {admm_plr.txt};
   	\addlegendentry{$\sigma_\mathrm{pm} = 0.5$}

  	\end{semilogyaxis}
 	\end{tikzpicture}
	\caption{RMSE vs Packet Loss Rate for the 30 bus IEEE test case with full
	PMU observability.}
	\label{Fig_ADMM2}
	\end{figure}
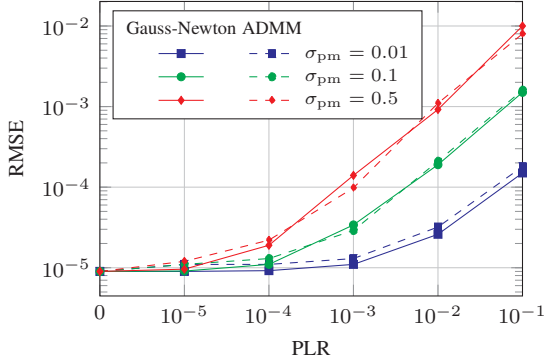

In Figure \ref{Fig_ADMM2}, we illustrate the RMSE performance of the ADMM-based scheme as a function of PLR. We consider the IEEE 30 bus test case scenario with PMU measurements only, with guaranteed observability, and corrupted with additive white Gaussian noise of standard deviation $\sigma=10^{-4}$. Whenever a measurement (packet) is dropped, it is replaced by a pseudo-measurement with higher standard deviation $\sigma_{\mathrm{pm}}$ (see Figure \ref{Fig_ADMM2}). The PLR of interest is in the range of $10^{-5}$ (for 5G URLLC service) and  $10^{-1}$ (for LTE without HARQ or RLC mechanisms in order to meet latency requirements, see Section IV-B). The performance of a centralized SE scheme based on the Gauss-Newton approach is also presented \cite{exposito}.

The performance attained by the distributed ADMM scheme for PLR of $10^{-5}$ (URLLC) is very close to that in the absence of packet losses. As expected, performance severely degrades by up to three orders of magnitude when PLR increases from $10^{-5}$ to $10^{-1}$ (i.e., in LTE range). The degradation is comparable for both the ADMM-based and Gauss-Newton approaches, however, ADMM operates in a distributed manner providing better scalability while preserving privacy.
	
\subsection{Distributed SE Methods: Performance vs Latency}

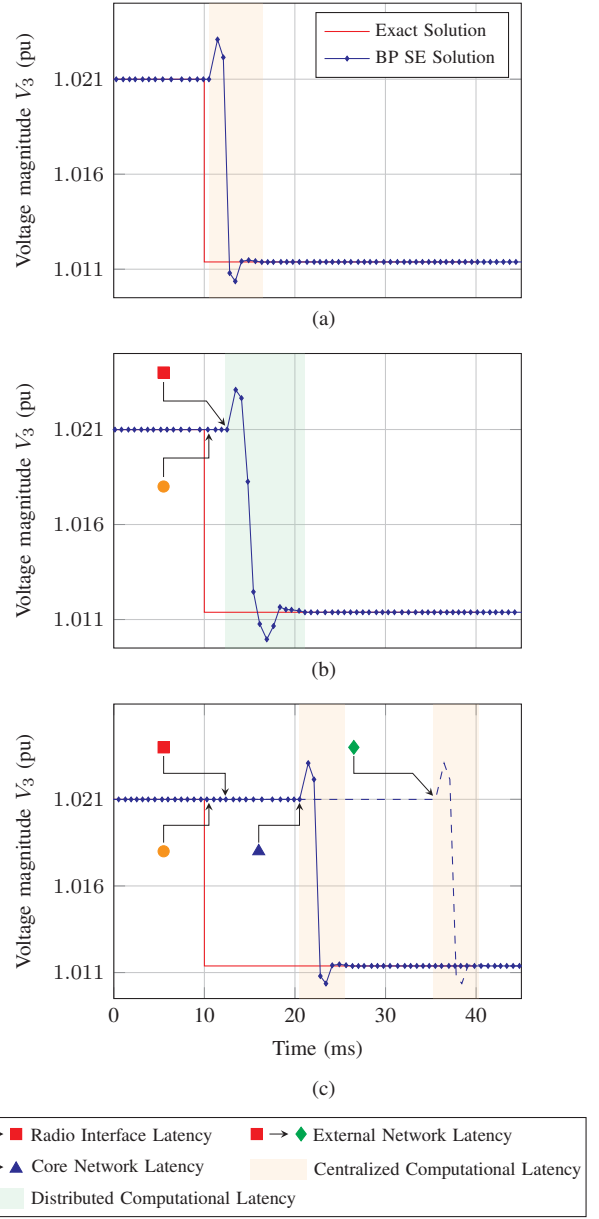
\begin{figure}[h]
	\centering
	\captionsetup[subfigure]{oneside,margin={1.7cm,0cm}}
	
	\begin{tabular}{@{}c@{}}
	\subfloat[]{\label{Fig_BP2_a}
	\centering	
	\begin{tikzpicture}	
	\begin{axis}[xmajorticks=false,width=7cm, height=5.5cm,,at={(0cm,0cm)},
   	y tick label style={{font=\footnotesize}, /pgf/number format/fixed,
    /pgf/number format/fixed zerofill,/pgf/number format/precision=3},
    legend style={legend pos=north east, font=\scriptsize, legend cell align=left},
   	legend entries={Exact Solution, BP SE Solution},
   	ylabel={Voltage magnitude $V_3$ (pu)},
   	label style={font=\footnotesize},
   	grid=major,
    ymin = 1.0095, ymax = 1.025,
    ytick={1.011, 1.016, 1.021},
   	xmin = 2.99, xmax = 3.035]

  	\fill [fill=orange!40, fill opacity=0.2] 
   	(axis cs:3.0005,1.001) rectangle (axis cs:3.0064,1.0285);
   	
    \addplot [red, no markers] coordinates {(2.9853269,1.02099501402936) 
    (3,1.02099501402936) (3,1.01138192767137) (3.03,1.01138192767137)}; 
    \addplot[mark=diamond*, mark repeat=1, mark size=0.9pt, blue] 
   	table [x={t}, y={V}] {bp_latency.txt};   	
   	\end{axis}
 	\end{tikzpicture}}	
	\end{tabular}\\[-0.5ex]
	
	\begin{tabular}{@{}c@{}}
	\subfloat[]{\label{Fig_BP2_b}
	\begin{tikzpicture}	
  	\begin{axis}[xmajorticks=false,width=7cm, height=5.5cm,
	y tick label style={{font=\footnotesize}, /pgf/number format/fixed,
    /pgf/number format/fixed zerofill,/pgf/number format/precision=3},
    ylabel={Voltage magnitude $V_3$ (pu)},
   	label style={font=\footnotesize},
   	grid=major,
    ymin = 1.0095, ymax = 1.025,
    ytick={1.011, 1.016, 1.021},
   	xmin = 2.99, xmax = 3.035]

	\fill [fill=green!40, fill opacity=0.2] 
   	(axis cs:3.0023,1.005) rectangle (axis cs:3.0111,1.03);
   	
    \addplot [red, no markers] coordinates {(2.9853269,1.02099501402936) 
    (3,1.02099501402936) (3,1.01138192767137) (3.03,1.01138192767137)}; 
    \addplot[mark=diamond*, mark repeat=1, mark size=0.9pt, blue] 
   	table [x={t1}, y={V1}] {bp_latency.txt};

	\draw[->, line width=0.1mm, >=stealth, black](axis cs:2.9955,1.0235)--
	(axis cs:2.9955,1.0225)--(axis cs:3.00023,1.0225)--(axis cs:3.0023,1.0212);
	\draw plot[mark=square*,mark options={color=red}, mark size=2.1pt] 
	coordinates {(axis cs:2.9955,1.024)};	

	\draw[->, line width=0.1mm, >=stealth, black](axis cs:2.9955,1.0185)--
	(axis cs:2.9955,1.0195)--(axis cs:3.0005,1.0195)--(axis cs:3.0005,1.0208);
	\draw plot[mark=*,mark options={color=orange}, mark size=2.1pt] 
	coordinates {(axis cs:2.9955,1.018)};				  	
   	\end{axis}
 	\end{tikzpicture}}
	\end{tabular}\\[-0.5ex]
	
	\begin{tabular}{@{}c@{}}
	\subfloat[]{\label{Fig_BP2_c}	
	\begin{tikzpicture}	
  	\begin{axis}[width=7cm, height=5.5cm,
   	x tick label style={{font=\footnotesize},/pgf/number format/.cd,
   	set thousands separator={},fixed},
   	y tick label style={{font=\footnotesize}, /pgf/number format/fixed,
    /pgf/number format/fixed zerofill,/pgf/number format/precision=3},
   	label style={font=\footnotesize},
   	ylabel={Voltage magnitude $V_3$ (pu)},
   	xlabel={Time (ms)},
   	grid=major,
    ymin = 1.0095, ymax = 1.0265,
    ytick={1.011, 1.016, 1.021},
   	xmin = 2.99, xmax = 3.035,
   	xtick={2.99, 3, 3.01, 3.02, 3.03},
   	xticklabels={0, 10, 20, 30, 40}]

	\fill [fill=orange!40, fill opacity=0.2] 
   	(axis cs:3.0105,1.001) rectangle (axis cs:3.0155,1.028); 
 
	\fill [fill=orange!40, fill opacity=0.2] 
   	(axis cs:3.02525,1.001) rectangle (axis cs:3.03025,1.028);  
   	
    \addplot [red, no markers] coordinates {(2.9853269,1.02099501402936) 
    (3,1.02099501402936) (3,1.01138192767137) (3.03,1.01138192767137)};
    \addplot[mark=diamond*, mark repeat=1, mark size=0.9pt, blue] 
   	table [x={t3}, y={V3}] {bp_latency.txt}; \label{cn}
   	\addplot[blue, dashed] 
   	table [x={t2}, y={V2}] {bp_latency.txt}; \label{en}

	\draw[->, line width=0.1mm, >=stealth, black](axis cs:2.9955,1.0235)--
	(axis cs:2.9955,1.0225)--(axis cs:3.0023,1.0225)--(axis cs:3.0023,1.0212);
	\draw plot[mark=square*,mark options={color=red}, mark size=2.1pt] 
	coordinates {(axis cs:2.9955,1.024)};

	\draw[->, line width=0.1mm, >=stealth, black](axis cs:2.9955,1.0185)--
	(axis cs:2.9955,1.0195)--(axis cs:3.0005,1.0195)--(axis cs:3.0005,1.0208);
	\draw plot[mark=*,mark options={color=orange}, mark size=2.1pt] 
	coordinates {(axis cs:2.9955,1.018)};
	
	\draw[->, line width=0.1mm, >=stealth, black](axis cs:3.006,1.0185)--
	(axis cs:3.006,1.0195)--(axis cs:3.0105,1.0195)--(axis cs:3.0105,1.0208);
	\draw plot[mark=triangle*,mark options={color=blue}, mark size=2.6pt] 
	coordinates {(axis cs:3.006,1.018)};	
					
	\draw[->, line width=0.1mm, >=stealth, black](axis cs:3.0165,1.0235)--
	(axis cs:3.0165,1.0225)--(axis cs:3.023,1.0225)--(axis cs:3.02525,1.0212);
	\draw plot[mark=diamond*,mark options={color=green}, mark size=2.6pt] 
	coordinates {(axis cs:3.0165,1.024)};		
   	\end{axis}	  	
 	\end{tikzpicture}}
	\end{tabular} \\
	\vspace{0,2cm}
	
	\begin{tabular}{@{}c@{}}
	\captionsetup[subfigure]{labelformat=empty}
	\subfloat[]{	
	\begin{tikzpicture}
 	\begin{axis}[ticks=none, width=10cm, height=2.9cm,
    ymin = 0, ymax = 12,
   	xmin = 0, xmax = 10]

	\draw plot[mark=*,mark options={color=orange}, mark size=2.1pt] 
	coordinates {(axis cs:0.3,10)};
	\draw[->, line width=0.1mm, >=stealth, black](axis cs:0.5,10)--
	(axis cs:0.8,10);
	\draw plot[mark=square*,mark options={color=red}, mark size=2.1pt] 
	coordinates {(axis cs:1.0,10)}
	node[pos=.5, text width=4.0cm, text=black, 
   	font=\scriptsize, align=center, fill opacity=1, yshift=6.7ex, xshift=14.2ex] 
   	{Radio Interface Latency}; 

	\draw plot[mark=square*,mark options={color=red}, mark size=2.1pt] 
	coordinates {(axis cs:4.8,10)};
	\draw[->, line width=0.1mm, >=stealth, black](axis cs:5.0,10)--
	(axis cs:5.3,10);
	\draw plot[mark=diamond*,mark options={color=green}, mark size=2.6pt] 
	coordinates {(axis cs:5.5,10)}
	node[pos=.5, text width=4.0cm, text=black, 
   	font=\scriptsize, align=center, fill opacity=1, yshift=6.7ex, xshift=38.67ex] 
   	{External Network Latency}; 

	\draw plot[mark=square*,mark options={color=red}, mark size=2.1pt] 
	coordinates {(axis cs:0.3,6)};
	\draw[->, line width=0.1mm, >=stealth, black](axis cs:0.5,6)--
	(axis cs:0.8,6);
	\draw plot[mark=triangle*,mark options={color=blue}, mark size=2.6pt] 
	coordinates {(axis cs:1.0,5.8)}
	node[pos=.5, text width=4.0cm, text=black, 
   	font=\scriptsize, align=center, fill opacity=1, yshift=4.1ex, xshift=13.9ex] 
   	{Core Network Latency}; 
   	  	
   	\fill [fill=orange!40, fill opacity=0.2] (axis cs:4.7,4.5) rectangle 
   	(axis cs:5.6,7)   	
	node[pos=.5, text width=4.0cm, text=black, 
   	font=\scriptsize, align=center, fill opacity=1, yshift=-0.15ex, xshift=14.2ex] 
   	{Centralized Computational Latency};
   	
   	 \fill [fill=green!40, fill opacity=0.2] (axis cs:0.21,1) rectangle 
   	(axis cs:1.12,3.5)   	
	node[pos=.5, text width=4.0cm, text=black, 
   	font=\scriptsize, align=center, fill opacity=1, yshift=-0.15ex, xshift=14.2ex] 
   	{Distributed Computational Latency};
   	\end{axis} 	
 	\end{tikzpicture}}
	\end{tabular}
	
	\caption{Latency performance of the BP-based SE for the IEEE 30 bus test 
	case where the BP-based SE: a) runs in a mode where communication latencies 
	are neglected, b) runs in a distributed mode accounting for radio 
	interface and MEC-to-MEC node latencies, and c) runs in the MCC node
	with radio interface and core network latency (\ref{cn}) or 
	external network latency (\ref{en}).}
	\label{Fig_BP2}
	\end{figure}

In Figure \ref{Fig_BP2}, we analyze latency of BP-based scheme in three different scenarios for the IEEE 30 bus test case. We analyzed the scenario where the system changes both generation and load values at the time instant $t = 10\,\mathrm{ms}$. As an example, we focus on the sudden bus voltage magnitude drop $V_3$. We assume measurements are obtained synchronously and immediately after the system change at $t = 10\,\mathrm{ms}$ using PMUs only, thus resulting in the linear SE model. The BP-based SE model runs the iterative message-passing algorithm continuously over time, with new measurements being integrated as they arrive. 

Figure \ref{Fig_BP2_a} illustrates the BP-based SE scenario which ignores communication latencies, while focusing only on computational latency measured from the time instant when the system acquired a new set of PMU measurements. The BP-based algorithm is able to provide fast response on the new state of the power system and steady state occurs after several BP iterations (note that figure markers denote BP outputs after each iteration). The computational latency can be additionally reduced if BP computations across variable/factor nodes in every iteration are parallelized.

Next, we assume PMUs deliver their measurements to MEC nodes through URLLC connection that introduces latency of $2\,\mathrm{ms}$. In addition, we consider the behavior of the distributed and asynchronous BP-based SE. More precisely, local MEC-SE modules run BP in a distributed fashion, where neighboring areas (MEC-SE modules) asynchronously exchange messages via X2 interfaces that introduce latency of $1\,\mathrm{ms}$. Figure \ref{Fig_BP2_b} shows that the BP algorithm requires more iterations to reach the steady state, due to delay of updates of inter-area BP messages, which results in computational latency increase. However, the BP algorithm is still able to track the system changes at the level of $\sim 10\,\mathrm{ms}$.      

Finally, we consider the scenario where PMU measurements are forwarded to the MCC node, where we observe two cases: i) MCC node resides within the mobile core network, or ii) MCC node is part of a data center in some external network. For the former case, typical core network latency of $10\,\mathrm{ms}$ are considered, while for the latter, additional latency of $20\,\mathrm{ms}$ towards external network is assumed. In both cases, the BP algorithm is implemented in the MCC node. In this framework, the SE model provides quasi real-time performance (see Figure \ref{fig_var}). We also note that if 4G LTE interface is used instead of 5G URLLC, the additional latency of $\sim 20-40\,\mathrm{ms}$ needs to be included. 

\section{Conclusions}

%Emerging 5G networks will offer a communication and computing platform suitable for real-time wide area monitoring and control of critical infrastructures. 

In this paper, we presented a convincing evidence that, in the forthcoming years, 5G technology will provide ideal arena for the development of future distributed Smart Grid services. These services will rely on massive and reliable acquisition of timely information from the system, in combination with large-scale computing and storage capabilities, providing highly responsive, robust and scalable monitoring and control solution for future Smart Grids.   

\section*{Acknowledgment}
This paper has received funding from the EU 7th Framework Programme for research, technological development and demonstration under grant agreement no. 607774.

\bibliographystyle{IEEEtran}
\bibliography{cite}

% Generated by IEEEtran.bst, version: 1.14 (2015/08/26)
\begin{thebibliography}{10}
\providecommand{\url}[1]{#1}
\csname url@samestyle\endcsname
\providecommand{\newblock}{\relax}
\providecommand{\bibinfo}[2]{#2}
\providecommand{\BIBentrySTDinterwordspacing}{\spaceskip=0pt\relax}
\providecommand{\BIBentryALTinterwordstretchfactor}{4}
\providecommand{\BIBentryALTinterwordspacing}{\spaceskip=\fontdimen2\font plus
\BIBentryALTinterwordstretchfactor\fontdimen3\font minus
  \fontdimen4\font\relax}
\providecommand{\BIBforeignlanguage}[2]{{%
\expandafter\ifx\csname l@#1\endcsname\relax
\typeout{** WARNING: IEEEtran.bst: No hyphenation pattern has been}%
\typeout{** loaded for the language `#1'. Using the pattern for}%
\typeout{** the default language instead.}%
\else
\language=\csname l@#1\endcsname
\fi
#2}}
\providecommand{\BIBdecl}{\relax}
\BIBdecl

\bibitem{exposito}
A.~Gomez-Exposito, A.~Abur, A.~de~la Villa~Jaen, and C.~Gomez-Quiles, ``A
  multilevel state estimation paradigm for smart grids,'' \emph{Proceedings of
  the IEEE}, vol.~99, no.~6, pp. 952--976, June 2011.

\bibitem{ghosh}
H.~Shariatmadari, R.~Ratasuk, S.~Iraji, A.~Laya, T.~Taleb, R.~Jantti, and
  A.~Ghosh, ``Machine-type communications: current status and future
  perspectives toward {5G} systems,'' \emph{IEEE Communications Magazine},
  vol.~53, no.~9, pp. 10--17, September 2015.

\bibitem{young}
Y.~C. Hu, M.~Patel, D.~Sabella, N.~Sprecher, and V.~Young, ``Mobile edge
  computing—a key technology towards {5G},'' \emph{ETSI White Paper},
  vol.~11, 2015.

\bibitem{yavuz}
A.~Rico-Alvarino, M.~Vajapeyam, H.~Xu, X.~Wang, Y.~Blankenship, J.~Bergman,
  T.~Tirronen, and E.~Yavuz, ``An overview of {3GPP} enhancements on machine to
  machine communications,'' \emph{IEEE Communications Magazine}, vol.~54,
  no.~6, pp. 14--21, June 2016.

\bibitem{addepalli}
F.~Bonomi, R.~Milito, J.~Zhu, and S.~Addepalli, ``Fog computing and its role in
  the internet of things,'' in \emph{Proceedings of the first edition of the
  {MCC} workshop on Mobile cloud computing}.\hskip 1em plus 0.5em minus
  0.4em\relax ACM, 2012, pp. 13--16.

\bibitem{turtinen}
A.~Maeder, A.~Ali, A.~Bedekar, A.~F. Cattoni, D.~Chandramouli,
  S.~Chandrashekar, L.~Du, M.~Hesse, C.~Sartori, and S.~Turtinen, ``A scalable
  and flexible radio access network architecture for fifth generation mobile
  networks,'' \emph{IEEE Communications Magazine}, vol.~54, no.~11, pp. 16--23,
  November 2016.

\bibitem{korres}
G.~N. Korres, ``A distributed multiarea state estimation,'' \emph{IEEE
  Transactions on Power Systems}, vol.~26, no.~1, pp. 73--84, Feb 2011.

\bibitem{minguez}
E.~Caro, A.~J. Conejo, and R.~Minguez, ``Decentralized state estimation and bad
  measurement identification: An efficient {L}agrangian relaxation approach,''
  \emph{IEEE Transactions on Power Systems}, vol.~26, no.~4, pp. 2500--2508,
  Nov 2011.

\bibitem{giannakis}
V.~Kekatos and G.~B. Giannakis, ``Distributed robust power system state
  estimation,'' \emph{IEEE Transactions on Power Systems}, vol.~28, no.~2, pp.
  1617--1626, May 2013.

\bibitem{haro}
J.~Matamoros, A.~Tsitsimelis, M.~Gregori, and C.~Anton-Haro, ``Multiarea state
  estimation with legacy and synchronized measurements,'' in \emph{2016 IEEE
  International Conference on Communications (ICC)}, May 2016, pp. 1--6.

\bibitem{zhu}
H.~Zhu and G.~B. Giannakis, ``Power system nonlinear state estimation using
  distributed semidefinite programming,'' \emph{IEEE Journal of Selected Topics
  in Signal Processing}, vol.~8, no.~6, pp. 1039--1050, Dec 2014.

\bibitem{kavcic}
Y.~Hu, A.~Kuh, T.~Yang, and A.~Kavcic, ``A belief propagation based power
  distribution system state estimator,'' \emph{IEEE Computational Intelligence
  Magazine}, vol.~6, no.~3, pp. 36--46, Aug 2011.

\bibitem{vukobratovic}
M.~Cosovic and D.~Vukobratovic, ``Distributed {G}auss-{N}ewton method for {AC}
  state estimation: A belief propagation approach,'' in \emph{2016 IEEE
  International Conference on Smart Grid Communications (SmartGridComm)}, Nov
  2016, pp. 643--649, extended version: https://arxiv.org/abs/1702.05781.

\bibitem{wiemann}
A.~Larmo, M.~Lindström, M.~Meyer, G.~Pelletier, J.~Torsner, and H.~Wiemann,
  ``The {LTE} link-layer design,'' \emph{IEEE Communications Magazine},
  vol.~47, no.~4, pp. 52--59, April 2009.

\bibitem{rupp}
M.~Laner, P.~Svoboda, P.~Romirer-Maierhofer, N.~Nikaein, F.~Ricciato, and
  M.~Rupp, ``A comparison between one-way delays in operating {HSPA} and {LTE}
  networks,'' in \emph{2012 10th International Symposium on Modeling and
  Optimization in Mobile, Ad Hoc and Wireless Networks (WiOpt)}, May 2012, pp.
  286--292.

\end{thebibliography}

\section*{Biography}
\vspace{-1.0cm}
\begin{IEEEbiographynophoto}{Mirsad Cosovic}
received Dipl.-Ing, and Mr.-Ing. Degrees in power electrical engineering from the University of Sarajevo, Faculty of Electrical Engineering, Bosnia and Herzegovina, in 2009 and 2013, respectively. Since December 2009, he was a Teaching Assistant at Faculty of Electrical Engineering of Sarajevo, and since October 2014, he is a Ph.D. Candidate as Marie Curie Early Stage Researcher at the University of Novi Sad, Faculty of Technical Sciences, Serbia.
\end{IEEEbiographynophoto}
\vspace{-0.7cm}
\begin{IEEEbiographynophoto}{Achilleas Tsitsimelis}
received Dipl.-Ing in Electrical and Computer Engineering in 2013 from National Technical University of Athens. During 2013 and 2014, he was working as a researcher at the Electrical and Computer Engineering School, NTUA-ICCS. Since 2014 he is a Marie Curie Early Stage Researcher at CTTC and he is pursuing his Ph.D in signal theory and communications department at UPC. His research interests include state estimation for the Smart Grid.
\end{IEEEbiographynophoto}
\vspace{-0.7cm}
\begin{IEEEbiographynophoto}{Dejan Vukobratovic}
received PhD degree in electrical engineering from the University of Novi Sad, Serbia, in 2008. Since 2009. he has been an Assistant Professor, and since 2014, an Associate Professor at the Department of Power, Electronics and Communication Engineering, University of Novi Sad. During 2009 and 2010, he was Marie Curie Intra-European Fellow at the University of Strathclyde, UK. His research interests include information and coding theory, wireless communications and signal processing.
\end{IEEEbiographynophoto}
\vspace{-0.7cm}
\begin{IEEEbiographynophoto}{Javier Matamoros}
holds a researcher position at CTTC. He received the M.Sc. degree in telecommunications and the Ph.D. degree in signal theory and communications from the Polytechnic University of Catalonia in 2005 and 2010, respectively. He has participated in several National and EC-funded projects (JUNTOS, NEWCOM\#, E2SG, EXALTED, ADVANTAGE). His primary research interests are on distributed optimization and machine learning applied to communications and Smart Grids.
\end{IEEEbiographynophoto}
\vspace{-0.7cm}
\begin{IEEEbiographynophoto}{Carles Ant\'on-Haro (M'99\textendash SM'03)}
holds a Ph.D. degree in telecommunications from UPC (cum-laude). In 1999, he joined Ericsson, where he participated in rollout projects of 2/3G networks. Currently, he is with the CTTC as a Director of R\&D Programs and Senior Researcher. His research interests include signal processing for communications (MIMO, WSN, 5G) and Smart Grids, optimization, estimation and control. He has published +30 articles in IEEE journals and +100 conference papers.
\end{IEEEbiographynophoto}

\vfill
\vfill
\end{document}